# On the Hierarchical Preconditioning of the PMCHWT Integral Equation on Simply and Multiply Connected Geometries

J. E. Ortiz Guzman[1], S. B. Adrian[1,2], R. Mitharwal[1], Y. Beghein[3], T. F. Eibert[2], K. Cools[4], and F. P. Andriulli[1]

[1]Institut Mines-Télécom / Télécom Bretagne, Technopôle Brest-Iroise, Brest, France
[2]Chair of High Frequency Engineering, Technical University of Munich, Munich, Germany
[3] Department of Information Technology, Ghent University, Ghent, Belgium
[4] The University of Nottingham, University Park, Nottingham, United Kingdom

*Abstract*—We present a hierarchical basis preconditioning strategy for the Poggio-Miller-Chang-Harrington-Wu-Tsai (PMCHWT) integral equation considering both simply and multiply connected geometries. To this end, we first consider the direct application of hierarchical basis preconditioners, developed for the Electric Field Integral Equation (EFIE), to the PMCHWT. It is notably found that, whereas for the EFIE a diagonal preconditioner can be used for obtaining the hierarchical basis scaling factors, this strategy is catastrophic in the case of the PMCHWT since it leads to a severely ill-conditioned PMCHWT system in the case of multiply connected geometries. We then proceed to a theoretical analysis of the effect of hierarchical bases on the PMCHWT operator for which we obtain the correct scaling factors and a provably effective preconditioner for both low frequencies and mesh refinements. Numerical results will corroborate the theory and show the effectiveness of our approach.

*Index Terms*—integral equation, PMCHWT, hierarchical basis, multiresolution, wavelet, preconditioner

## I. Introduction

In the frequency domain, the Poggio-Miller-Chang-Harrington-Wu-Tsai (PMCHWT) integral equation is used to solve scattering problems involving dielectric bodies. The PMCHWT operator comprises the Electric Field Integral Equation (EFIE) operator and the Magnetic Field Integral Equation (MFIE) operator and, thereby, it inherits some of the properties of these operators. In particular, if the frequency $f$ or the average edge length $h$ of the mesh decreases, then the condition number of the system matrix of the discretized PMCHWT operator grows. These effects are often referred to as low-frequency and dense-discretization breakdown, respectively [1].

Classically, the low-frequency breakdown has been cured by using loop-tree or loop-star preconditioners [2]. More recently, a Calderón identity based preconditioner has been presented for curing also the dense-discretization breakdown [3]–[5]. This method, however, requires the use of dual elements defined on the barycentric refinement of the mesh. On the other hand, hierarchical basis preconditioners have been used in the past to cure the low-frequency and dense-discretization breakdown of the EFIE without using barycentric refinements [6], [7].

This work was supported in part by the Agence Nationale de la Recherche under the Project FASTEEG-ANR-12-JS09-0010 and in part by the COMINLabs Excellence Laboratory under the project SABRE, grant reference ANR-10-LABX-07-0.

For this reason it would be useful to have a hierarchical basis strategy to precondition the PMCHWT, both in frequency and in discretization, without the need to go on the dual mesh. The extension of an EFIE hierarchical basis strategy to the PMCHWT, however, is not straightforward due to the fundamentally different way in which the two equations act on the global loops of the structure, the so called harmonic subspace.

The contribution of this paper is twofold: (i) We present a theoretical analysis of the low-frequency properties of the PMCHWT operator at low frequencies for both simply and multiply connected geometries. The analysis will show why a direct extension of EFIE strategies would not work for the PMCHWT and will clarify both conditioning properties and solution scalings of the classical and preconditioned equation. (ii) We present a hierarchical basis preconditioning strategy that solves the low-frequency and the dense-discretization breakdown of the PMCHWT for both simply and multiply connected geometries.

For the sake of completeness the reader should notice that the conference contribution [8] reported numerical results of the application of a hierarchical basis to the PMCHWT. Although the approach used by the authors of [8] remains undefined in their paper, their numerical experiments are limited to simply connected geometries (as for dielectrics) and thus they are not relevant for the theoretical findings presented here.

This paper is organized as follows: Section 2 sets the notation and introduces background material, in Section 3 we derive our theory and propose a new preconditioner for the PMCHWT. Numerical results are presented and discussed in Section 4 that both corroborate the theory and show the practical applicability of the new scheme.

## II. Notation and Background

We consider a polyhedral domain $\Omega_i \subset \mathbb{R}^3$ with intrinsic impedance $\eta_i$ and boundary $\Gamma = \partial \Omega_i$, which can be simply or multiply connected. The exterior domain $\Omega_o = \Omega_i^c \setminus \partial \Omega_i$ has the intrinsic impedance $\eta_o$. The EFIE operator $\mathcal{T}^\kappa = i\kappa \mathcal{T}_A^\kappa + 1/(i\kappa)\mathcal{T}_\Phi^\kappa$, is constituted by the vector potential $\mathcal{T}_A^\kappa f = \hat{n} \times \int_\Gamma G(r, r') f(r') dS(r')$ and the scalar potential $\mathcal{T}_\Phi^\kappa f = -\hat{n} \times \nabla_\Gamma \int_\Gamma G(r, r') \nabla'_\Gamma \cdot f(r') dS(r')$, and the MFIE operator is $\mathcal{K}^\kappa f = -\hat{n} \times \int_\Gamma \nabla_\Gamma G(r, r') \times f(r') dS(r')$, where $\hat{n}$ is the outward going normal to the surface $\Gamma$, the wavenumber $\kappa$ is $k_\mathrm{i}$ or $k_\mathrm{o}$ associated with the domain $\Omega_\mathrm{i}$ or $\Omega_\mathrm{o}$, and $G(\boldsymbol{r},\boldsymbol{r}') = \mathrm{e}^{\mathrm{i}\kappa|\boldsymbol{r}-\boldsymbol{r}'|}/(4\pi|\boldsymbol{r}-\boldsymbol{r}'|)$ is the free-space Green's function. A time-harmonic electromagnetic wave $(\boldsymbol{E}^\mathrm{i},\boldsymbol{H}^\mathrm{i})$ is impinging on $\Omega_\mathrm{i}$. Note that a time dependency $\mathrm{e}^{-\mathrm{i}\omega t}$ is suppressed throughout this paper. The PMCHWT integral equation reads

$$\begin{bmatrix} \mathcal{T}^{k_\mathrm{i}}/\eta_\mathrm{i} + \mathcal{T}^{k_\mathrm{o}}/\eta_\mathrm{o} & -(\mathcal{K}^{k_\mathrm{i}}+\mathcal{K}^{k_\mathrm{o}}) \\ (\mathcal{K}^{k_\mathrm{i}}+\mathcal{K}^{k_\mathrm{o}}) & \eta_\mathrm{i}\mathcal{T}^{k_\mathrm{i}}+\eta_\mathrm{o}\mathcal{T}^{k_\mathrm{o}} \end{bmatrix} \begin{bmatrix} \boldsymbol{M} \\ \boldsymbol{J} \end{bmatrix} = \begin{bmatrix} -\hat{\boldsymbol{n}}\times\boldsymbol{H}^\mathrm{i} \\ -\hat{\boldsymbol{n}}\times\boldsymbol{E}^\mathrm{i} \end{bmatrix},$$

for all $\boldsymbol{r} \in \Gamma$. It relates the magnetic and electric surface current densities $\boldsymbol{M}$ and $\boldsymbol{J}$, defined on $\Gamma$, to the incident fields.

To obtain $\boldsymbol{M}$ and $\boldsymbol{J}$, the surface $\Gamma$ is triangulated and, via a Galerkin approach, the currents are approximated using the Rao-Wilton-Glisson (RWG) functions $\boldsymbol{f}_n$ as source and the rotated RWGs $\hat{\boldsymbol{n}}\times\boldsymbol{f}_n$ are used as testing functions so that the currents are approximated as $\boldsymbol{M} = \sum_{n=1}^{N}[m]_n \boldsymbol{f}_n$ and $\boldsymbol{J} = \sum_{n=1}^{N}[j]_n \boldsymbol{f}_n$. The RWGs are normalized such that the flux through their defining edges equals 1. Thus, the linear system to solve reads

$$Z\begin{bmatrix}m\\i\end{bmatrix} := \begin{bmatrix} \mathsf{T}^{k_\mathrm{i}}/\eta_\mathrm{i}+\mathsf{T}^{k_\mathrm{o}}/\eta_\mathrm{o} & -(\mathsf{K}^{k_\mathrm{i}}+\mathsf{K}^{k_\mathrm{o}}) \\ \mathsf{K}^{k_\mathrm{i}}+\mathsf{K}^{k_\mathrm{o}} & \eta_\mathrm{i}\mathsf{T}^{k_\mathrm{i}}+\eta_\mathrm{o}\mathsf{T}^{k_\mathrm{o}} \end{bmatrix}\begin{bmatrix}m\\j\end{bmatrix} = \begin{bmatrix}h\\e\end{bmatrix}, \quad (1)$$

where the matrices are $[\mathsf{T}^\kappa]_{mn} = (\hat{\boldsymbol{n}}\times\boldsymbol{f}_m,\mathcal{T}^\kappa\boldsymbol{f}_n)_{L^2(\Gamma)}$ and $[\mathsf{K}^\kappa]_{mn} = (\hat{\boldsymbol{n}}\times\boldsymbol{f}_m,\mathcal{K}^\kappa\boldsymbol{f}_n)_{L^2(\Gamma)}$ using the $L^2(\Gamma)$-duality pairing. The right-hand side vectors are defined as $[h]_m = (\hat{\boldsymbol{n}}\times\boldsymbol{f}_m,-\hat{\boldsymbol{n}}\times\boldsymbol{H}^\mathrm{i})_{L^2(\Gamma)}$ and $[e]_m = (\hat{\boldsymbol{n}}\times\boldsymbol{f}_m,-\hat{\boldsymbol{n}}\times\boldsymbol{E}^\mathrm{i}(\boldsymbol{r}))_{L^2(\Gamma)}$.

Any hierarchical basis $\boldsymbol{H}_n$ for the EFIE is a quasi-Helmholtz decomposition consisting of solenoidal $\boldsymbol{H}_n^\Lambda$, non-solenoidal $\boldsymbol{H}_n^\Sigma$, and quasi-harmonic functions $\boldsymbol{H}_n^\mathrm{qH}$. These functions are defined as linear combinations of RWG functions and we have $\boldsymbol{J} \approx \sum_{n=1}^{N}[j]_n\boldsymbol{f}_n = \sum_{n=1}^{N_\Lambda}[j_\Lambda]_n\boldsymbol{H}_n^\Lambda + \sum_{n=1}^{N_\mathrm{qH}}[j_\mathrm{qH}]_n\boldsymbol{H}_n^\mathrm{qH} + \sum_{n=1}^{N_\Sigma}[j_\Sigma]_n\boldsymbol{H}_n^\Sigma$. The reader should recall that the quasi-harmonic functions $\boldsymbol{H}_n^\mathrm{qH}$ are related to the global cycles of the structure, that is, they are present only when $\Gamma$ is multiply connected to represent the quasi-harmonic Helmholtz subspace [9]. In fact, only when the global loops are added, we have $N = N_\Lambda + N_\mathrm{qH} + N_\Sigma$ with $N_\mathrm{qH} = 2g$, where $g$ is the genus of the geometry. Since the functions $\boldsymbol{H}_n^\Lambda$, $\boldsymbol{H}_n^\mathrm{qH}$, and $\boldsymbol{H}_n^\Sigma$ are defined as linear combinations of RWG functions, we can define transformation matrices $\mathsf{H}_\Lambda$, $\mathsf{H}_\mathrm{qH}$, and $\mathsf{H}_\Sigma$ such that $j = \mathsf{H}_\Lambda j_\Lambda + \mathsf{H}_\mathrm{qH} j_\mathrm{qH} + \mathsf{H}_\Sigma j_\Sigma$. The analysis of this work applies to any hierarchical basis which can precondition the EFIE, that is, it yields a condition number that grows at most $O(\log(1/h)^2)$. In this work, we use the matrices $\mathsf{H}_\Lambda$, $\mathsf{H}_\mathrm{qH}$, and $\mathsf{H}_\Sigma$ obtained from the hierarchical basis defined in [6], for which after defining the diagonal matrices $[\mathsf{D}_\Lambda]_{nn} = 2^{l_\Lambda(n)/2}$, $[\mathsf{D}_\Sigma]_{nn} = 2^{-l_\Sigma(n)/2}$ (where the function $l_\Lambda(n)$, $n \in \{0\dots,N_\Lambda\}$, returns the level on which $\boldsymbol{H}_n^\Lambda$ is defined), and $\mathsf{M}_k = [\mathsf{H}_\Lambda \mathsf{D}_\Lambda/\sqrt{k} \quad \mathsf{H}_\mathrm{qH}/\sqrt{k} \quad \mathsf{H}_\Sigma \mathsf{D}_\Sigma \sqrt{k}]$, we have the conditioning property

$$\mathrm{cond}(\mathsf{M}_k^\mathrm{T}\mathsf{T}^k\mathsf{M}_k) \lesssim \log^2(1/h^2). \quad (2)$$

In other words, for any hierarchical basis preconditioner for the EFIE satisfying condition (2), the theoretical developments obtained in this work will hold.

## III. Hierarchical Preconditioners for the PMCHWT

The diagonal blocks of the PMCHWT operator contain the EFIE operator. The most intuitive idea for preconditioning the PMCHWT would then be to use the same EFIE strategy: perform the same change of basis for both electric and magnetic currents (as delineated in the previous section) and then proceed with a diagonal preconditioning. This idea would surely render well-conditioned diagonal blocks of the PMCHWT operator. Unfortunately, however, such a strategy would have catastrophic effects due to the off-diagonal blocks of the PMCHWT operator as will be shown next.

To see and solve this problem, we will perform a complete frequency analysis of the PMCHWT operator and associated solutions. To this purpose, we will use the hierarchical basis transformation without rescaling $\mathsf{M} = [\mathrm{diag}(\mathsf{M}_1,\mathsf{M}_1)]$ with $\mathsf{M}_1 = [\mathsf{H}_\Lambda \quad \mathsf{H}_\mathrm{qH} \quad \mathsf{H}_\Sigma]$. This is possible since a hierarchical basis is in particular also a quasi-Helmholtz decomposition.

In order to proceed, we need to use the frequency analysis in [2] and extend it to the case of non-simply connected geometries. In [2], in fact, it is observed that the magnetostatic field produced by a loop is curl-free. This implies that the discretization of the $\mathcal{K}^k$ operator scales as $O(k^2)$ whenever the source is solenoidal (a local or a global loop) and the testing is a local loop, or the source is a local loop and the testing is solenoidal (a local or a global loop). For a further explanation the reader is referred to [2]. From this it follows that the hierarchical matrix blocks $\mathsf{H}_\Lambda^\mathrm{T}\mathsf{K}^k\mathsf{H}_\Lambda$, $\mathsf{H}_\Lambda^\mathrm{T}\mathsf{K}^k\mathsf{H}_\mathrm{qH}$, and $\mathsf{H}_\mathrm{qH}^\mathrm{T}\mathsf{K}^k\mathsf{H}_\Lambda$ will scale as $O(k^2)$. It remains to be studied the case in which both source and testing functions are global loops (i.e., they belongs to the quasi-harmonic subspace). We prove here the following result that will be used both here and in the $h$-conditioning analysis that will follow:

**Proposition 1.** *The matrix block $\mathsf{H}_\mathrm{qH}^\mathrm{T}\mathsf{K}^k\mathsf{H}_\mathrm{qH}$ scales as $O(1)$ with respect to $k$ and has a conditioning which is both frequency and $h$ independent.*

*Proof:* First it should be remembered that the harmonic Helmholtz subspace $H = \mathrm{span}\{\boldsymbol{h}_i\}$ is spanned by two $g$-dimensional, orthogonal subspaces, the poloidal and the toroidal loops [10]. Since these two subspaces are in the null-spaces of the operators $\pm\mathcal{I}/2 - \mathcal{K}^0$, respectively, [10], we have $\mathcal{K}^0 = \mathcal{I}/2$ for toroidal loops and $\mathcal{K}^0 = -\mathcal{I}/2$ for poloidal loops, where $\mathcal{I}$ is the identity operator. If the RWG functions could span $H$ exactly, the above property would immediately prove the hypothesis. However, the RWG functions cannot span $H$ exactly, instead they provide functions that are solenoidal, but not necessarily harmonic. This is equivalent to saying that linear combinations of RWGs result in global loops $\boldsymbol{H}_i^\mathrm{qH} = \boldsymbol{h}_i + \boldsymbol{s}_i$, with $\boldsymbol{s}_i$ a solenoidal and non-irrotational function (i.e., a local loop). However it should be noted that $(\hat{\boldsymbol{n}}\times\boldsymbol{h}_j,\mathcal{K}^0\boldsymbol{s}_i)_{L^2} = (\hat{\boldsymbol{n}}\times\boldsymbol{s}_j,\mathcal{K}^0\boldsymbol{h}_i)_{L^2} = (\hat{\boldsymbol{n}}\times\boldsymbol{s}_j,\mathcal{K}^0\boldsymbol{s}_i)_{L^2} = 0$, so that $(\hat{\boldsymbol{n}}\times(\boldsymbol{h}_j+\boldsymbol{s}_j),\mathcal{K}^0(\boldsymbol{h}_i+\boldsymbol{s}_i))_{L^2} = (\hat{\boldsymbol{n}}\times\boldsymbol{h}_j,\mathcal{K}^0\boldsymbol{h}_i)_{L^2}$ from which the proposition is proved. ∎

The above proposition completes the analysis of the off-diagonal blocks of the PMCHWT. The analysis of the diagonal blocks and of the right-hand side (plane wave-excitation) directly follows from the EFIE [11]. Altogether, the following scalings for the entire PMCHWT equation, valid for both simply and multiply connected geometries are obtained (note that for the sake of brevity we write $k$ instead of $O(k)$) :

$$M_{H_\Lambda}^T ZM\left[M^{-1}[j;m]\right] = M^T[h;e] =$$

$$\begin{array}{c} H_\Lambda^T \\ H_{qH}^T \\ H_\Sigma^T \\ H_\Lambda^T \\ H_{qH}^T \\ H_\Sigma^T \end{array} \underbrace{\begin{bmatrix} k & k & k & k^2 & k^2 & 1 \\ k & k & k & k^2 & 1 & 1 \\ k & k & k^{-1} & 1 & 1 & 1 \\ k^2 & k^2 & 1 & k & k & k \\ k^2 & 1 & 1 & k & k & k \\ 1 & 1 & 1 & k & k & k^{-1} \end{bmatrix}}_{H_\Lambda \ H_{qH} \ H_\Sigma \ H_\Lambda \ H_{qH} \ H_\Sigma} \begin{bmatrix} 1 \\ k \\ 1 \\ 1 \\ k \\ k \end{bmatrix} \begin{array}{l} \}M_1^{-1}j \\ \\ \}M_1^{-1}m \end{array} = \begin{bmatrix} k \\ k \\ 1 \\ k \\ k \\ 1 \end{bmatrix} \begin{array}{l} \}M_1^T h \\ \\ \}M_1^T e \end{array}$$

(3)

where the scaling of the solution has been obtained after performing a block analysis based on the Sherman-Morrison formulas [12]. The scalings in equation (3) clearly show why the idea of using the same preconditioning strategy adopted for the EFIE would be catastrophic for the PMCHWT. In fact, a diagonal preconditioner, by definition, would render $O(1)$ the frequency scaling of elements (2,2) and (5,5), but as a consequence the off-diagonal blocks associated to the harmonic subspace (elements (2,5) and (5,2)) would scale as $O(1/k)$ resulting in a catastrophic growth of the condition number.

Given the analysis above, we propose a hierarchical preconditioned PMCHWT in the form of

$$L^T ZR y = L^T [h;e], \quad \text{with } j = Ry. \quad (4)$$

The (potentially different) left and right preconditioning matrices we propose are defined as $L = [\text{diag}(L_k, L_k)]$, $R = [\text{diag}(R_k R_k)]$, and with $L_k = \begin{bmatrix} H_\Lambda D_\Lambda/\sqrt{k} & H_{qH}\sqrt{\beta} & iH_\Sigma D_\Sigma \sqrt{k} \end{bmatrix}$ and $R_k = \begin{bmatrix} H_\Lambda D_\Lambda/\sqrt{k} & H_{qH}\sqrt{\alpha} & iH_\Sigma D_\Sigma \sqrt{k} \end{bmatrix}$. The imaginary scaling of the non-solenoidal functions homogenizes the sign of the overall operator in the static limit [13]. Moreover, since a special problem has been identified by the previous analysis in the scaling of the harmonic subspace, we will study the effect of different scalings via the constants $\alpha$ and $\beta$.

With these definitions we obtain the following frequency scaling for the preconditioned equation: $L^T ZR =$

$$\begin{bmatrix} 1 & \sqrt{\alpha}\sqrt{k} & k & k & k\sqrt{k}\sqrt{\alpha} & 1 \\ \sqrt{\beta}\sqrt{k} & \sqrt{\alpha\beta}k & \sqrt{\beta}k\sqrt{k} & \sqrt{\beta}k\sqrt{k} & \sqrt{\alpha\beta} & \sqrt{\beta}\sqrt{k} \\ k & k\sqrt{k}\sqrt{\alpha} & 1 & 1 & \sqrt{k}\sqrt{\alpha} & k \\ k & k\sqrt{k}\sqrt{\alpha} & 1 & 1 & \sqrt{k}\sqrt{\alpha} & k \\ \sqrt{\beta}k\sqrt{k} & \sqrt{\alpha\beta} & \sqrt{\beta}\sqrt{k} & \sqrt{\beta}\sqrt{k} & \sqrt{\alpha\beta}k & \sqrt{\beta}k\sqrt{k} \\ 1 & \sqrt{k}\sqrt{\alpha} & k & k & k\sqrt{k}\sqrt{\alpha} & 1 \end{bmatrix}$$

(5)

while for the current and the right-hand-side the scalings are $R^{-1}[j;m]^T = \begin{bmatrix} \sqrt{k} & k/\sqrt{\alpha} & \sqrt{k} & \sqrt{k} & k/\sqrt{\alpha} & \sqrt{k} \end{bmatrix}^T$ and $L[h;e]^T = \begin{bmatrix} \sqrt{k} & k\sqrt{\beta} & \sqrt{k} & \sqrt{k} & k\sqrt{\beta} & \sqrt{k} \end{bmatrix}^T$. Several choices of $\alpha$ and $\beta$ are possible, but the following constraints arise by the scalings above: (i) $\alpha = 1/\beta$ in order to avoid a frequency breakdown in the off-diagonal block (elements (2,5) and (5,2). (ii) $\alpha$ and $\beta$ can grow at most as $O(1/k)$ to avoid breakdowns. Two notable choices of $\alpha$ and $\beta$ arise from this analysis:(a) $\alpha = 1/\beta = k$ which provides an homogeneous frequency scaling of $O(\sqrt{k})$ for both solution and right-hand-side and (b) $\alpha = \beta = 1$ which provide a symmetric frequency

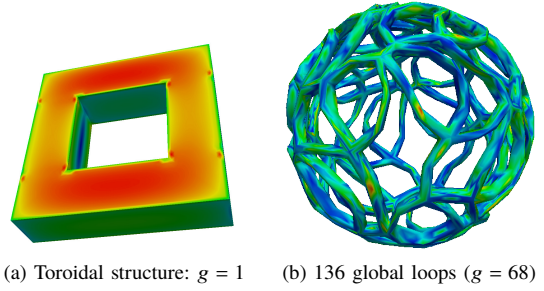

(a) Toroidal structure: $g = 1$  (b) 136 global loops ($g = 68$)

Fig. 1. Test geometries with real part of the electric current density.

scaling in the preconditioned matrix. We opt here for (b) since in this case $L = R$, which reduces the preconditioning storage. The other choice, however, could be equally exploited.

From (5) with $\alpha = \beta = 1$, it follows immediately that $\text{cond}(L^T ZR) = O(1), k \rightarrow 0$. Moreover, the $h$-dependent conditioning of the preconditioned equation is also uniformly bounded up to logarithmic terms. In fact, the diagonal blocks have a bounded conditioning as it follows from the initial hypothesis (2). The off-diagonal blocks are instead compact when either source or testing space is different from the harmonic subspace. In the case in which both source and testing functions are in the harmonic subspace, the $h$-uniform conditioning is ensured by the structure in (5) and by Proposition 1. Summarizing, the preconditioned PMCHWT we proposed will have a bounded condition number both in frequency and in discretization. Although the theoretical analysis is rigorous only for smooth surfaces, numerical practice in hierarchical schemes has shown that these methods are robust even for non-uniformly meshed and non-smooth geometries.

## IV. Numerical Results

To verify the frequency performance of our new scheme, we compared our new formulation with a standard loop-star preconditioner and with a hierarchical scheme followed by a naive diagonal preconditioner. For this test, we used a structure with two global loops shown in Fig. 1a with permittivity $\varepsilon_r = 3$ and with a maximum diameter of 2.8 m. We used a plane wave excitation and the conjugate gradient squared (CGS) method (other Krylov methods show a similar behavior). Fig. 2 shows that the number of iterations needed by the CGS solver to obtain a relative residual of $10^{-6}$ is independent from the frequency for both Loop-Star and the formulation presented in this work. However, the new formulation needs fewer iterations as expected by the fact that loop-star/tree schemes have a derivative behavior [14]. It should also be noted that a hierarchical scheme followed by a naive diagonal preconditioner does not deliver a frequency-independent number of iterations. This confirms the theory we have developed in the previous section. To verify the dense-discretization stability, we compared our new formulation again with a standard loop-star preconditioner and used a plane wave excitation with frequency 1 MHz. Fig. 3 shows the number of iterations needed by the CGS solver to obtain a relative residual of $10^{-6}$. We can see that the number of iterations needed by our new formulation is almost constant as predicted by theory while the derivative nature of the loop-star preconditioner results in a $h$-growing condition number.

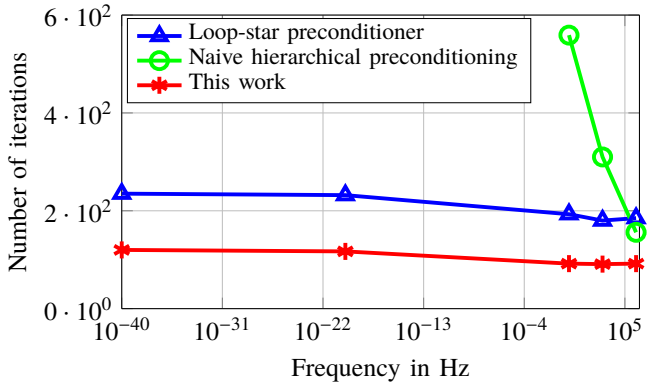

Fig. 2. Toroidal structure: the number of iterations as a function of the frequency.

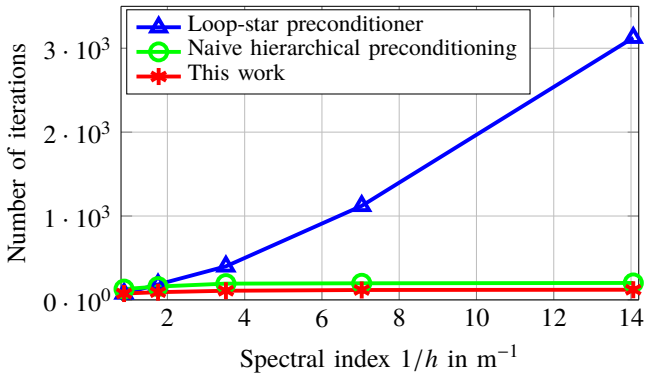

Fig. 3. Toroidal structure: the number of iterations as a function of the spectral index $1/h$.

For $h$-refinement, naive diagonal preconditioning also yields a $O(\log(1/h)^2)$ conditioning which, however, deteriorates with respect to the new scheme by a constant depending on the frequency.

To verify the effectiveness of our approach in the presence of a larger harmonic subspace, we tested our new formulation on the multiply connected structure in Fig. 1b containing 136 global loops and 2.7 m of diameter. A plane wave excitation is used with frequency 1 MHz. The results are shown in Tab. I. Again there is a substantial advantage with respect to loop-star techniques even if only one dyadic refinement step was used as in this case. The stability on a harmonic subspace is thus also confirmed numerically.

| Formulation | Iterations | Time |
|---|---|---|
| Loop-star preconditioner | 5396 | 52h 1'22" |
| Naive hierarchical preconditioning | 18318 | »100h |
| This work | 2642 | 21h 6'5" |

TABLE I
136 GLOBAL LOOPS STRUCTURE: THE NUMBER OF ITERATIONS FOR THE DIFFERENT FORMULATIONS WITH SOLVER TOLERANCE $10^{-6}$.

Finally, to verify the accuracy, we compared the results obtained by our preconditioner with an analytic solution. Fig. 4a and Fig. 4b shows a good agreement of the radar cross section (RCS) for a sphere with radius 1 m and relative permittivity

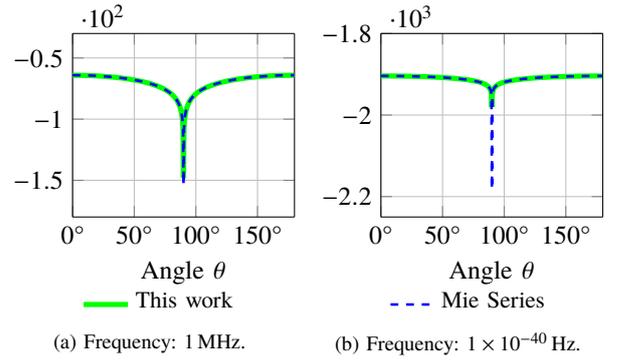

(a) Frequency: 1 MHz.  (b) Frequency: $1 \times 10^{-40}$ Hz.

Fig. 4. Sphere: Radar cross section in dBsm.

$\varepsilon_\mathrm{r} = 3$ and plane wave excitation. In both cases the error is $-22.6\,\mathrm{dB}$